\documentclass{article}

\usepackage{arxiv}

\usepackage[utf8]{inputenc} 
\usepackage[T1]{fontenc}  
\usepackage{hyperref}    
\usepackage{url}      
\usepackage{booktabs}    
\usepackage{amsfonts}    
\usepackage{nicefrac}    
\usepackage{microtype}   
\usepackage{lipsum}		
\usepackage{graphicx}

\usepackage{subcaption}
\usepackage{float}

\PassOptionsToPackage{%
	backend=bibtex8,bibencoding=ascii,%
	language=auto,%
    style=authoryear-comp, 
    bibstyle=authoryear,dashed=false, 
    sorting=nyt, 
    maxbibnames=10, 
    natbib=true 
}{biblatex}
    \usepackage{biblatex}
\usepackage{doi}

\title{Data driven weather forecasts trained and initialised directly from observations}

\author{{Anthony McNally}
    \And {Christian Lessig}
    \And {Peter Lean}
    \And {Eulalie Boucher}
    \And {Mihai Alexe}
    \And {Ewan Pinnington}
    \And {Matthew Chantry}
    \And {Simon Lang}
    \And {Chris Burrows}
    \And {Marcin Chrust}
    \And {Florian Pinault}
    \And {Ethel Villeneuve}
    \And {Niels Bormann}
    \And {Sean Healy} 
    \And
    European Centre for Medium-Range Weather Forecasts (ECMWF)}

\begin{document}

\maketitle

\begin{abstract}
 Skilful Machine Learned (ML) weather forecasts have challenged conventional approaches to numerical weather prediction (NWP), demonstrating competitive performance compared to traditional physics-based approaches. Existing data-driven systems have been trained to forecast future weather by learning from long historical records of past weather, typically provided by reanalyses such as ECMWF's ERA5. These datasets have been made freely available to the wider research community, including the commercial sector, which has been a major factor in the rapid rise of ML forecast systems and the impressive levels of accuracy they have achieved. However, both historical reanalyses used for training and real-time analyses used for initial conditions are produced by data assimilation, essentially an optimal blending of observations with a traditional physics-based forecast model. As such, many ML forecast systems have an implicit, unknown and unquantified dependence on the physics-based models they seek to challenge. Here we propose a new and radical approach to weather forecasting, by training a neural network to predict future weather purely from historical observations with no dependence on a physics-based model or reanalysis datasets. We use raw observations (level-1) to initialise a model of the atmosphere (in observation space) learned directly from the observations themselves. Forecasts of crucial weather-related parameters (such as surface temperature and wind) are obtained by predicting weather parameter observations (such as SYNOP surface data) at future times and arbitrary locations. We present preliminary results on forecasting observations 12-hours into the future. These already demonstrate successful learning of the time evolution of the physical processes captured in real observations. We argue that this new approach, by staying purely in observation space avoids many of the challenges of traditional data assimilation, can exploit a wider range of observations and is readily expanded to simultaneous forecasting of the full Earth system (atmosphere, land, ocean and atmospheric composition).

\end{abstract}


\section*{The current use of observations in NWP}

The many millions of meteorological observations measured each day require highly sophisticated data assimilation (DA) systems to transform the raw measurements made at irregular times and locations into a best estimate of the state of the atmosphere on a regular grid and into the variables required to initialise physics-based forecast models. Data assimilation is based upon very sound theoretical principles and performs this task by combining different sources of information in an optimal way, respecting the uncertainty in each element.  The European Centre for Medium-range Weather Forecasts (ECMWF) and others have been extremely successful in the development of advanced global data assimilation systems (e.g. 4D-Var, \citet{rabier20004dvar}) that exploit a large amount of information from the observations and generate highly skilful initial conditions. These DA systems, in addition to initialising real-time operational forecast models, have been used to create state of the art historical weather records such as ERA5 (\citet{hersbach2020era5}) used extensively for climate research and, more recently, training ML forecast models. However, DA is an extremely challenging task because forecast models demand initialisation on a fine spatial grid over the entire globe (e.g. 9~km for the ECMWF Integrated Forecast System) and many vertical levels, even in regions where no observations are available. The challenge is further compounded as physics-based models require initialisation in physical meteorological variables (e.g. temperature, humidity, and wind) while the overwhelming majority of weather observations come from passive satellite sensors which measure thermal infrared and microwave radiance signals emanating from broad vertical layers of the atmosphere. In data assimilation systems, the extraction of temperature, humidity or wind requires a detailed understanding of complex radiative transfer processes in the atmosphere, including a prior knowledge of the current atmospheric state. These downward viewing satellites, by their nature, also cannot provide information on fine vertical scales (the existence of a so-called null space, is discussed in \citet{rodgers1976retrieval}, atmospheric structures to which the radiances are effectively blind). To meet the demands of forecast model initialisation, the data assimilation system must very carefully blend the satellite radiances and other observations with gridded fine-scale information from a previous forecast trajectory, known as a background. This operation requires a highly exacting characterisation of large, complex and often poorly known observation and background error covariances. In practice, and particularly in a real-time operational DA context, many compromises have to be made in the exploitation of observations. One of the most significant is that large fractions of the global observing system are not actually used for operational NWP. Observations typically need to be heavily thinned in both space and time, discarded in certain meteorological conditions when uncertainties become inconsistent with our error covariances, or even discarded completely when the physics of the measurement is extremely difficult to relate to the model variables.

\section*{Does machine learning offer an alternative? }

Much of the data assimilation complexity described above stems from the fact that physics-based forecast models demand extremely accurate initial conditions for all meteorological variables on a fine spatial grid. The problem is effectively ill-posed, in that there is insufficient information contained in the observations to constrain all parameters of this 3D grid. The strategy that has traditionally been employed by NWP centres to produce accurate weather forecasts is to build models which represent the real atmosphere as comprehensively as possible, explicitly describing a myriad of physical processes and interactions at every grid point over the entire globe, from the surface up to the mesosphere \citep{bauer2015quiet}. Over recent decades, we have witnessed impressive accuracy gains in global NWP as ever more powerful supercomputers have allowed physics-based models to operate at ever higher spatial resolutions, and the representation of unresolved processes has improved. And as model resolution has increased, so has the burden placed upon data assimilation systems to provide the initial conditions with the required accuracy and detail. 

However, more recently it has been demonstrated that ML systems operating with far fewer variables than physics-based models and on significantly coarser spatial grids are capable of producing highly skilful medium-range forecasts of important weather parameters \citep{bi2023accurate, lam2023learning, price2023gencast, lang2024aifs}. Given the dependence of these systems upon reanalysis datasets (for training) and real-time analyses (for initialisation), both produced by data assimilation (with the aforementioned challenges and limitations), we ask the intriguing question if ML forecast models could learn and be initialised directly from observations. This would obviate the need for DA systems to map raw observations (e.g. satellite radiances) to a fine grid of unmeasured physical variables (e.g. temperature) dictated by the requirements of the physics-based model and one could instead directly use the information in the observations to produce a forecast. If training of a ML model directly from observations is successful, large fractions of the data assimilation process described above could be completely reconsidered. For real-time applications like operational NWP (both physics-based or the current generation of analysis-based data driven models) this could allow forecasts to be produced in a significantly timelier manner, effectively as soon as observations are available. But crucially, observation-based prediction in observation space could also be more accurate. A model learned directly from observations might be more primitive than comprehensive physics-based systems, but if it allows the exploitation of more observations (more than are used in operational analyses and reanalyses) with fewer approximations and avoids the null-space problem, the end result has the potential to be more accurate. 

In the prototype experiments described here, a machine learning algorithm has been trained to predict future observations from long time series of historical observations. A key challenge is to learn the spatial and temporal correlations that exist within a given observation type, but also between different measurement systems, and encapsulate these within the internal latent space of the machine learning model. Learning the relationships between different observations (both the complementarities and the redundancies) from the historical data allows the network at inference time to predict future observations. However, predicting future values of satellite radiances at different microwave and infrared wavelengths would not constitute a useful weather forecast. To be useful, the network must be able to predict future weather parameter observations such as SYNOP surface measurements and radiosondes measurements in the upper atmosphere, by learning the correlations between these sparse in-situ data and the dense global satellite radiance observations. The spatial and temporal relationships between weather station data and the satellite radiances, once learned from real observation locations, must then be generalised to arbitrary (user defined) locations to allow weather parameter prediction wherever required.

\section*{Observation training data}

Our first prototype experiments have used a limited set of observations as detailed in Table~\ref{tab:table_1}. The table highlights the instruments, variables and period of each dataset used and specifies whether the data is used prognostically or diagnostically. Prognostic observations are passed both to the network as inputs and are predicted as an output, whilst diagnostic observations are only predicted but are not part of the inputs (and are thus derived from other input sources). 

\begin{table}[h!]
    \centering
    \begin{tabular}{|l|l|l|l|l|}
    \hline
        \textbf{Instrument} & \textbf{Satellite} & \textbf{Period} & \textbf{Variables} & \textbf{Use} \\
        \hline
        
AMSUA &	METOP-B	& 2013- present & 15 channels & Prognostic \\
\hline
ATMS	&NPP, NOAA-20	&2012 - present&	22 channels	&Prognostic\\
\hline

IASI&	METOP-B &	2013 - present	&17 channels &	Prognostic\\ \hline
AVHRR	&METOP-B &	2013 - present&	Visible channel	&Prognostic\\ \hline
SYNOPS&	N/A &	1979 – present &	ps, t2m, rh2m, u10, v10	&Diagnostic \\ \hline

    \end{tabular}

    \caption{A description of the current set of observations in the training database. A glossary of satellite instrument acronyms is contained in appendix A.}
    \label{tab:table_1}
\end{table}

\begin{figure}[h!]
\centering
\begin{subfigure}[t]{0.45\textwidth}
        \centering
        \includegraphics[width=\textwidth]{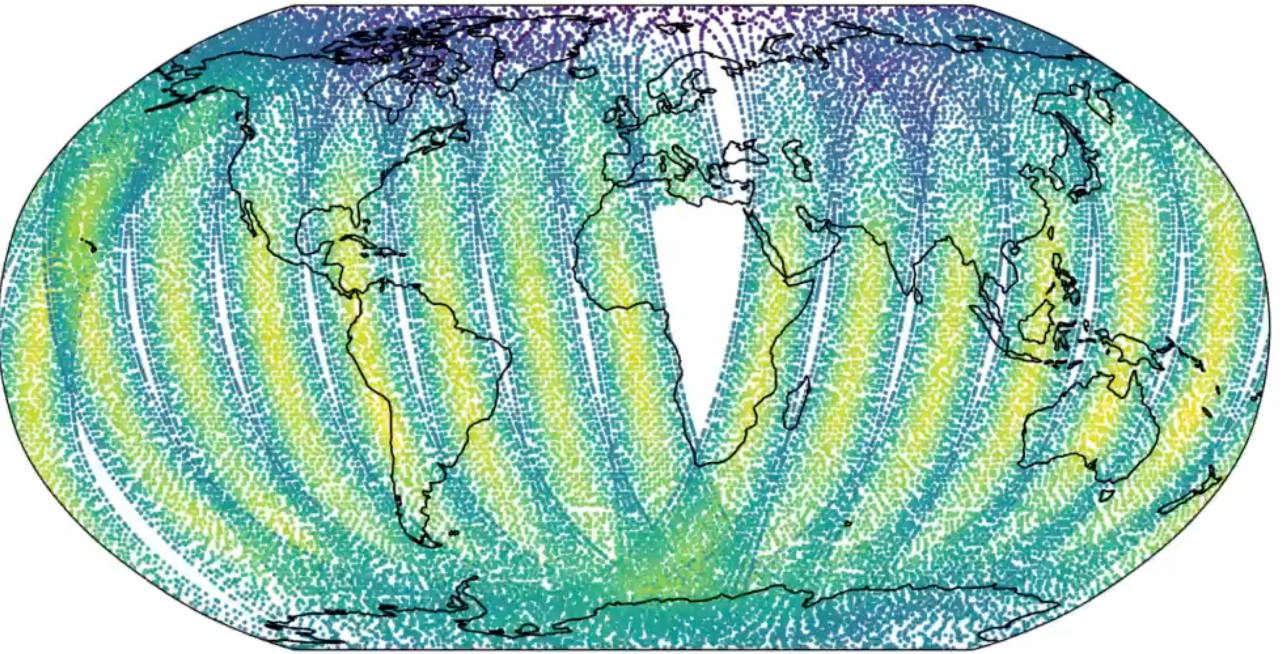}
        \caption{ATMS microwave radiances}
    \end{subfigure}
\hfill
    \begin{subfigure}[t]{0.45\textwidth}
        \centering
        \includegraphics[width=\textwidth]{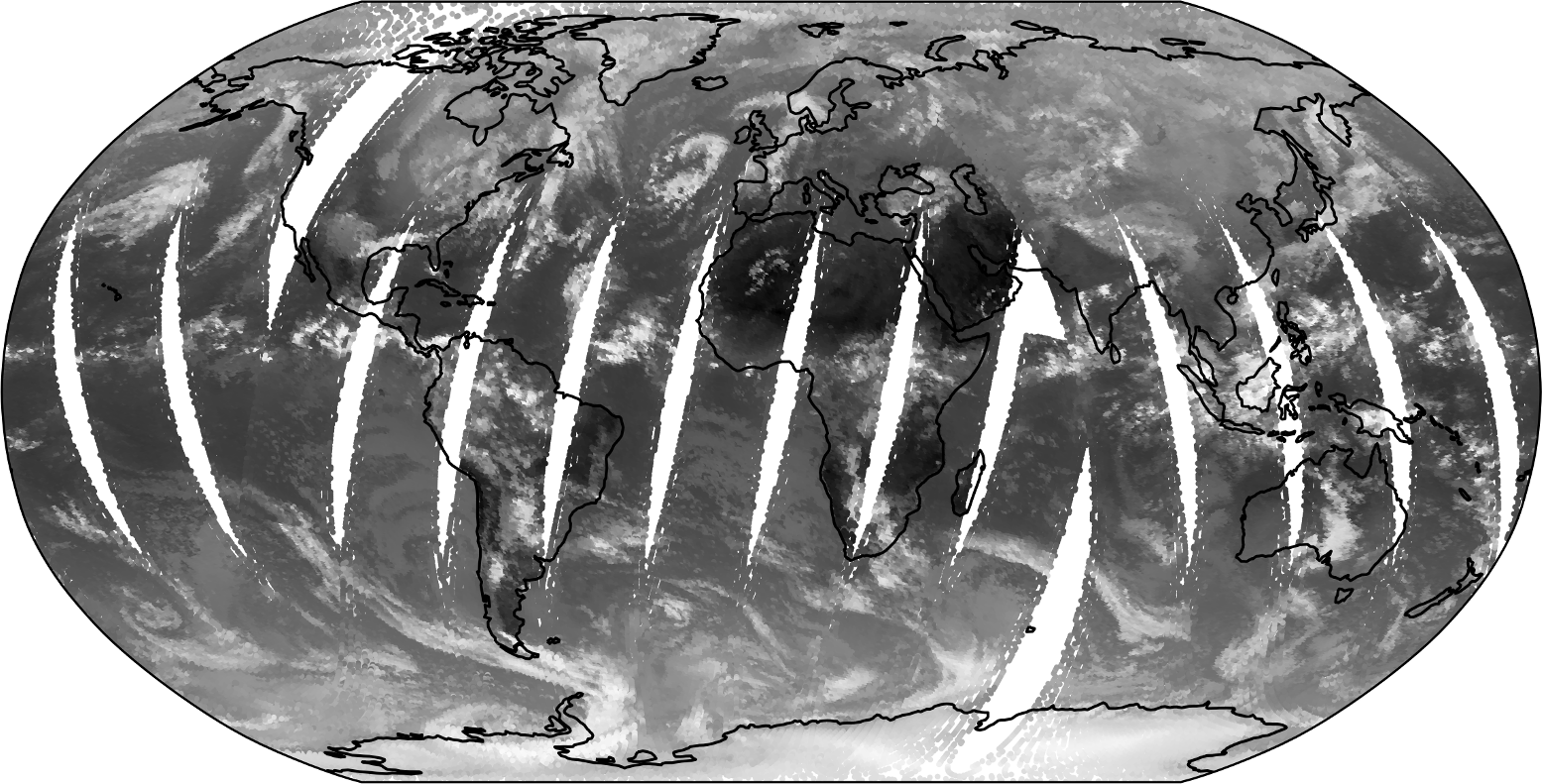}
        \caption{IASI infrared radiances}
        
    \end{subfigure}
    \vskip\baselineskip 
    
    \begin{subfigure}[t]{0.45\textwidth}
        \hfill 
        \includegraphics[width=\textwidth]{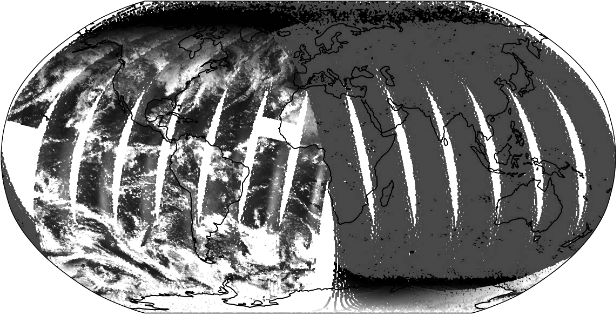}
        \caption{AVHRR visible reflectances}
        
    \end{subfigure}
\hfill    
    \begin{subfigure}[t]{0.45\textwidth}
        \centering
        \includegraphics[width=\textwidth]{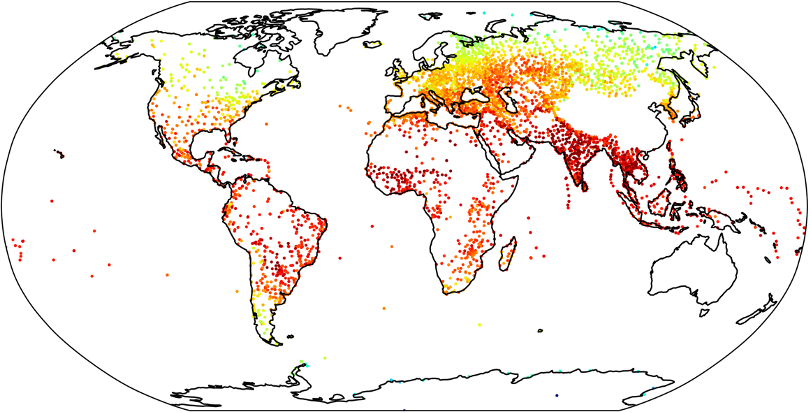}
        \caption{SYNOP surface measurements of 2m temperature and 10m wind}
    \end{subfigure}

\caption{Typical examples of the data coverage provided by ATMS microwave radiances (a), IASI infrared radiances (b), AVHRR visible reflectances (c) and SYNOP surface measurements of 2m temperature and 10m wind (d) within a 12-hour window.}
\label{fig:fig_1}
\end{figure}

Passive microwave radiance measurements from ATMS and AMSU sensors (see glossary in appendix A for instrument acronym meanings) onboard low Earth orbiting (LEO) satellites provide deep layer information on temperature (from channels around 50~GHz) and on water vapour (from channels around 183~GHz). Microwave window channels also provide information on surface temperature, albeit potentially contaminated by highly variable surface emissivity values. Similar deep layer temperature and humidity information is provided by passive infrared radiances measured by the IASI instrument, also carried by LEO satellites. Emissivity over land and ocean is less variable and closer to unity at infrared wavelengths, providing more direct information on surface temperature. Infrared radiances display an acute (and highly nonlinear) sensitivity to the presence of clouds. While this sensitivity significantly limits the use of infrared data in cloudy regions by conventional data assimilation systems, it is not an obstacle for a machine learning system. On the contrary, the spatial and temporal patterns in the cloudy infrared data provide the network with invaluable information on the trajectory of weather systems as they move around the globe. Similar information on the movement of weather systems is provided by AVHRR visible imagery, the reflected solar illumination providing better thermal contrast for cloud evolutions over cold higher-latitude surfaces compared to the infrared and free from the potentially interfering effects of humidity. Figure~\ref{fig:fig_1}~(a)-(c) shows that all of the satellite observations provide near global spatial coverage within a 12-hour window and frequent temporal sampling when carried on multiple LEO satellites. Also, in situ observations of surface temperature and wind from SYNOP stations are included in the training. Coverage of these data (see Figure~\ref{fig:fig_1}~(d)) can be very sparse (e.g. over oceans and uninhabited land areas) and even over populated areas it can be rather inhomogeneous. However, these weather parameter observations provide an important strategic resource to the training process, allowing the network to forecast weather parameters, by exploiting correlations learned between these and satellite radiance observations. 


\section*{Prototype neural network}

\begin{figure}[h!]
    \centering
    \includegraphics[trim={8cm 3cm 0 0}, clip, width=0.9\linewidth]{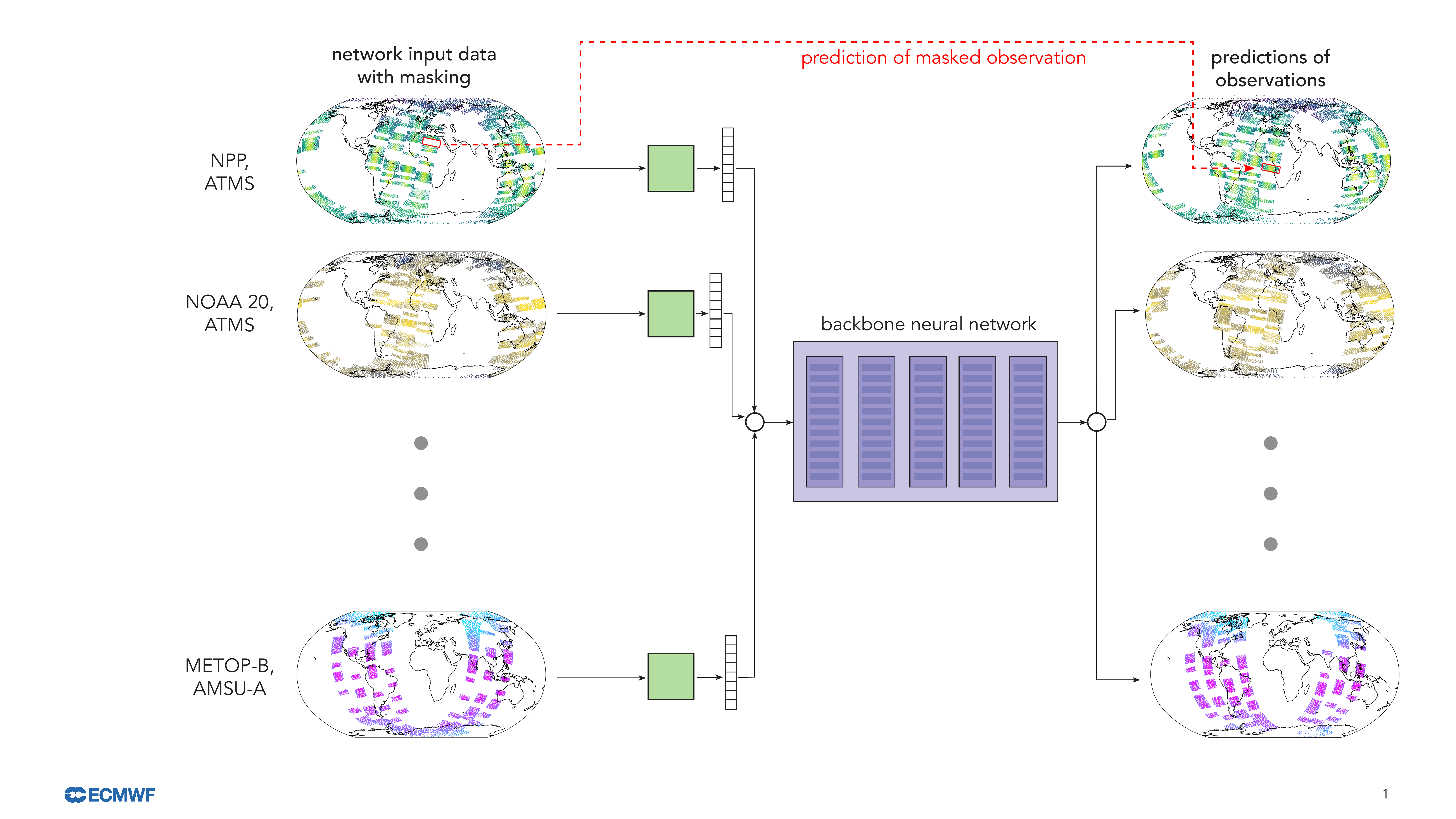}
    \caption{A schematic representation of the transformer neural network used in the prototype experiments. On the left we see portions of the satellite observations being randomly masked, and during training the network being challenged to predict the values of the masked data. }
    \label{fig:fig_2}
\end{figure}

The prediction of future observations from historical observations should be reasonably agnostic to the choice of network architecture, as long as the training data is sufficiently rich. For our prototype experiments, we use a transformer-based neural network depicted schematically in Figure~\ref{fig:fig_2}. The network takes as input information subdivided into tokens that represent data in a small space-time neighbourhood. The tabular form of the data are mapped through learnable embedding networks to a latent vector representation. For each observation, all channels are treated together and embedded together with their associated location as well as observation time. For radiosondes, weather parameters on different standard pressure levels are treated in a similar way. The set of vectors from all neighbourhoods and all observations is the input to the transformer ‘backbone’ neural network, in which the data interacts through the transformer’s attention mechanism \citep{vaswani2017attention}. The attention mechanism is a key component of transformer models, learning how each element of the representation relates to the other elements. Here it learns how spatial, temporal and observation-specific information are related. 

During training, randomly selected parts of the data are masked with only the space-time coordinates of the observation being fed to the embedding networks. The pre-training task is to predict the corresponding withheld observed values. The network is then fine-tuned to make 12-hour predictions, essentially using observations in one 12-hour assimilation time window as input and predicting the observations that would be obtained in the next 12-hour window. This selection of time window is somewhat arbitrary but chosen here to allow the exploitation of pre-existing infrastructure supporting the ECMWF operational data assimilation system. It also allows some comparisons to be made between the network predictions and observations in the 12-hour 4D-Var.

\section*{Results of the prototype experiments}

\begin{figure}[h]
    \centering
    \begin{subfigure}[t]{0.45\textwidth}
        \centering
        \includegraphics[width=\textwidth]{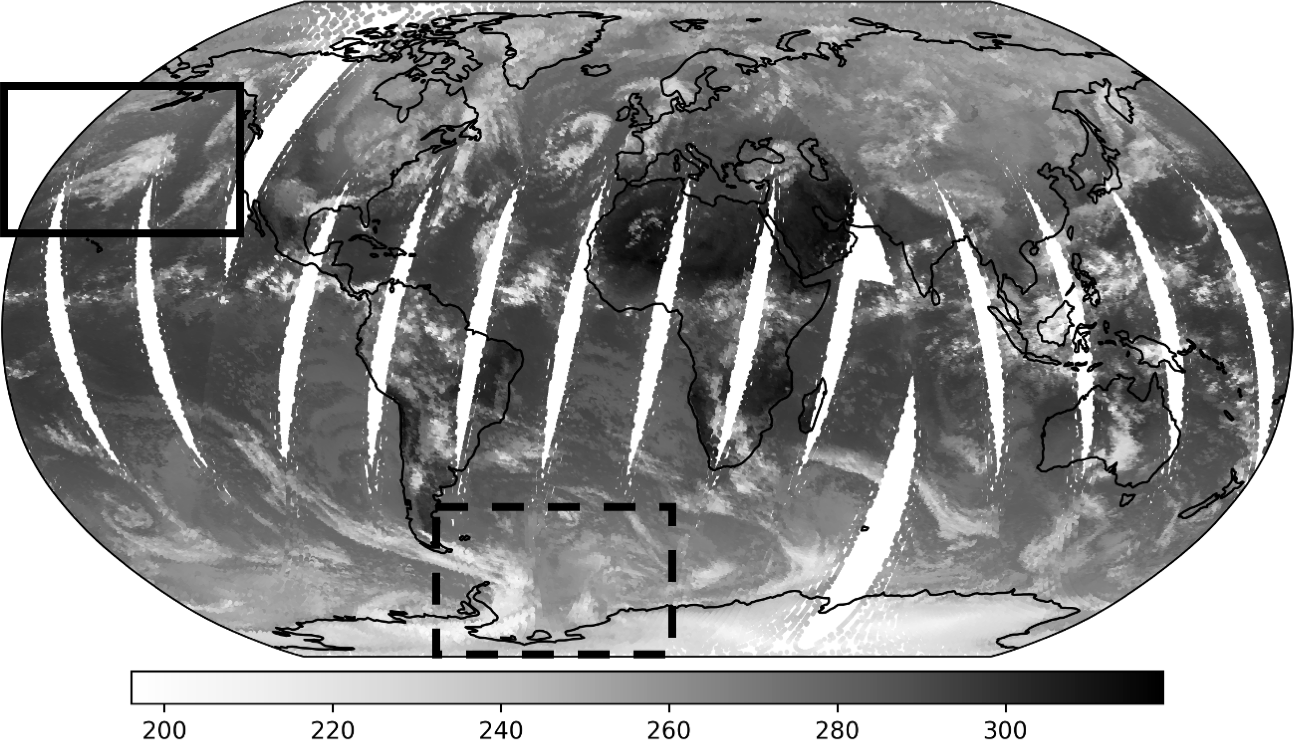}
        \caption{Input IASI channel 921 radiances}
        
    \end{subfigure}
    \hfill
    \begin{subfigure}[t]{0.45\textwidth}
        \centering
        \includegraphics[width=\textwidth]{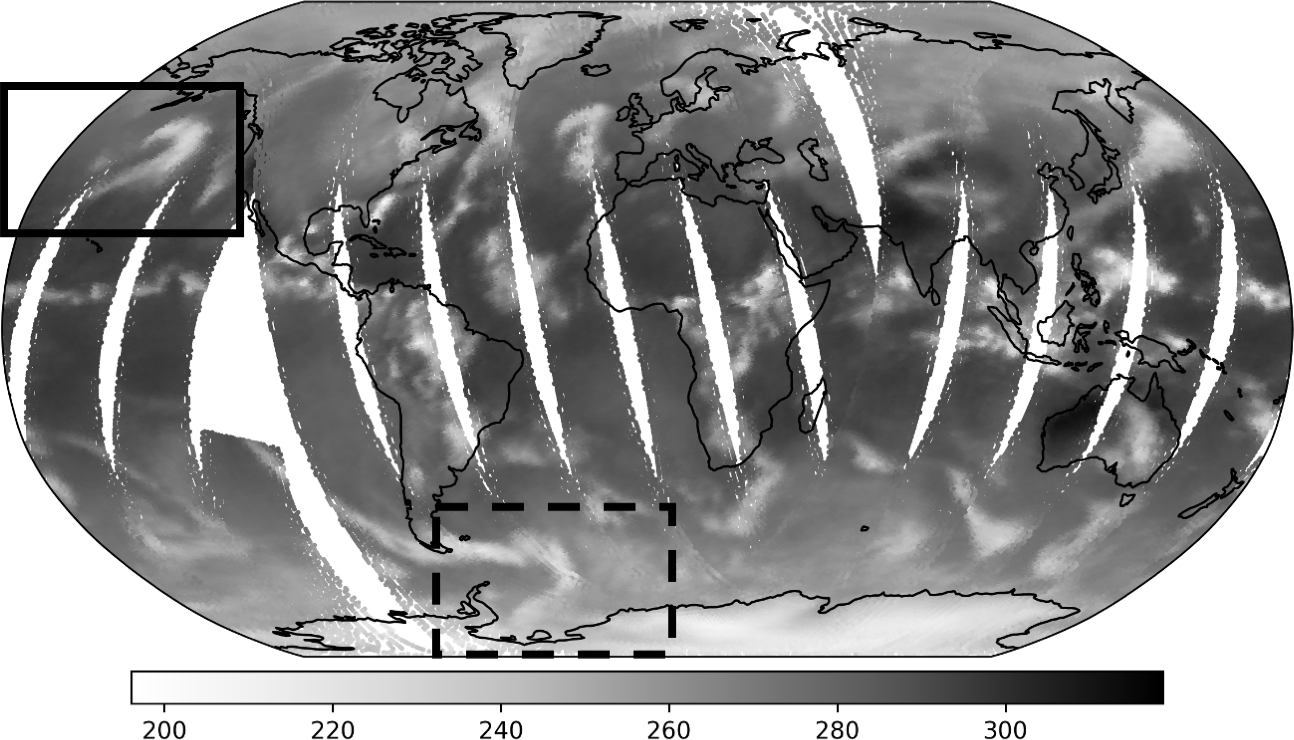}
        \caption{Predicted IASI channel 921 radiances}
        
    \end{subfigure}
    \vskip\baselineskip 
    \hspace*{\fill} 
    \begin{subfigure}[t]{0.45\textwidth}
        \hfill 
        \includegraphics[width=\textwidth]{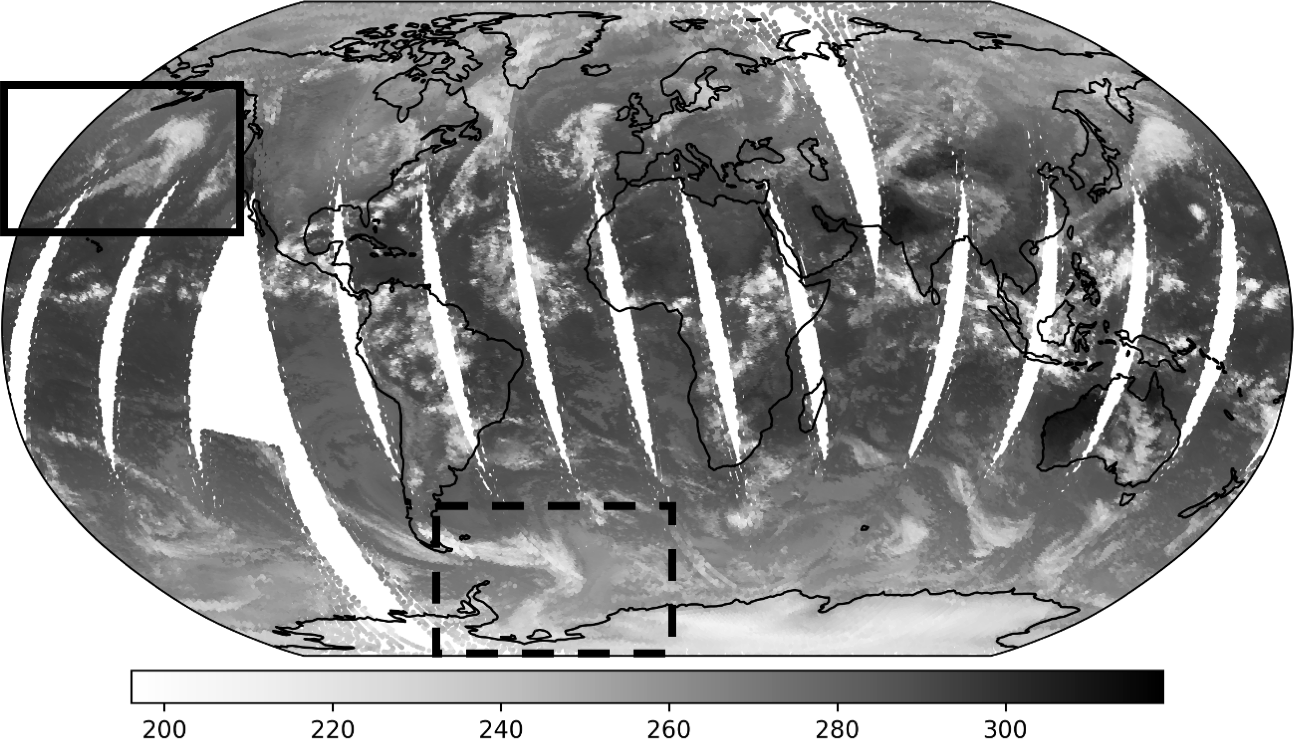}
        \caption{Target IASI channel 921 radiances}
        
    \end{subfigure}

\caption{ An example of the network predicting 12-hours of IASI window channel radiances (channel 921 around 10 microns) for a case on October 17th 2022 (00z) using the measurements from the previous 12-hour window as input. In this IASI channel, light shades correspond to cold features (e.g. clouds and the Antarctic continent) and dark shades to warm features (e.g. the satellite viewing warm surfaces unobscured by clouds).  }
\label{fig:fig_3}
\end{figure}

An example of the successful prediction of future observations is shown qualitatively in Figure~\ref{fig:fig_3} for IASI infrared radiances. The future IASI radiances predicted (b) by the network are clearly physically plausible in comparison to real measurements (c). The predicted values are a smoothed representation of the real data, but otherwise capture the large-scale patterns and indeed many of the finer scale cloud patterns in a very reasonable way. What is also evident in Figure~\ref{fig:fig_3} is that the network is correctly predicting the advection of weather systems (for example the easterly movement over 12 hours of the Southern Hemisphere frontal cloud pattern contained the dashed box), but also more complex atmospheric evolutions such as those shown within the solid box over the North Pacific. Here we see more than simple advection with the formation of a hook-shaped cloud feature that was not in the initial conditions (observations) 12 hours previously. 


Over equatorial Africa there is some evidence that the predictions successfully capture changes to the deep convection that occur over 12 hours in the real observations. The realism of the forecasted satellite radiance evolutions is similarly evident in the microwave radiance and visible reflectance satellite data (not shown). 

\begin{figure}[h]
    \centering
    \begin{subfigure}[t]{0.45\textwidth}
        \centering
        \includegraphics[width=\textwidth]{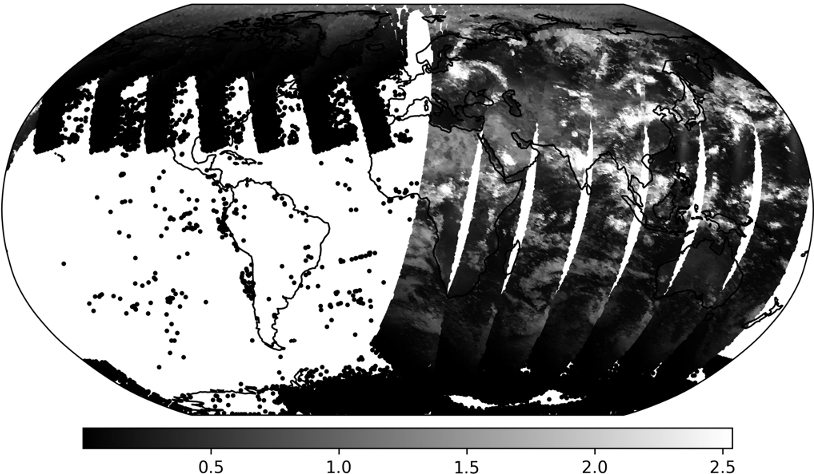}
        \caption{Input AVHRR visible reflectances}
        
    \end{subfigure}
    \hfill
    \begin{subfigure}[t]{0.45\textwidth}
        \centering
        \includegraphics[width=\textwidth]{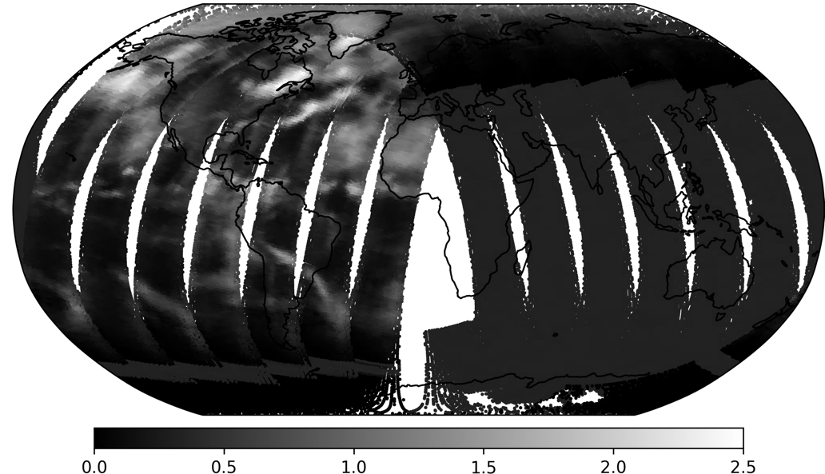}
        \caption{Predicted AVHRR visible reflectances}
        
    \end{subfigure}
    \vskip\baselineskip 
    \hspace*{\fill} 
    \begin{subfigure}[t]{0.45\textwidth}
        \hfill 
        \includegraphics[width=\textwidth]{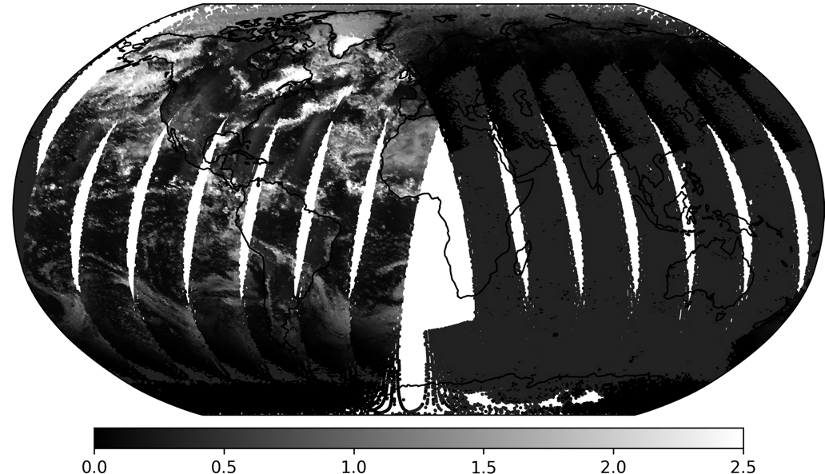}
        \caption{Target AVHRR visible reflectances}
        
    \end{subfigure}
    \caption{ Example of input (a), predicted (b) and target (c) METOP-B AVHRR visible reflectances in the next 12-hour window on October 17th 2022 (00z), along with the target truth. Note the accurate prediction of visible cloud features in the western hemisphere despite this hemisphere being in darkness for the input window and having no AVHRR information.  }
    \label{fig:fig_4}
\end{figure}

In the results of the prototype experiments we also see evidence that the network is able to make skillful predictions that exploit the correlations between different observing systems contained in the historical training data. Figure~\ref{fig:fig_4} shows 12 hours of real AVHRR data provided as input to the prediction, where the western hemisphere is mostly in darkness and the reflectances are zero (apart from at high summer northern latitudes where there is solar illumination and some sporadic calibration anomalies). Yet, in the subsequent 12-hour window the network successfully predicts very plausible reflectances for the western hemisphere which is now illuminated by the sun. This prediction skill cannot have originated from the AVHRR reflectances in the western hemisphere used as input 12 hours previously (as these were zero) so the network must have drawn upon information from other observing systems. The most likely explanation is that the network has learned the very high correlation between clouds causing bright solar reflections in the AVHRR visible data and clouds causing strong absorption signals in the IASI infrared measurements (similar to those seen in Figure~\ref{fig:fig_3}).

\begin{figure}[h!]
    \centering
    \includegraphics[width=0.98\linewidth]{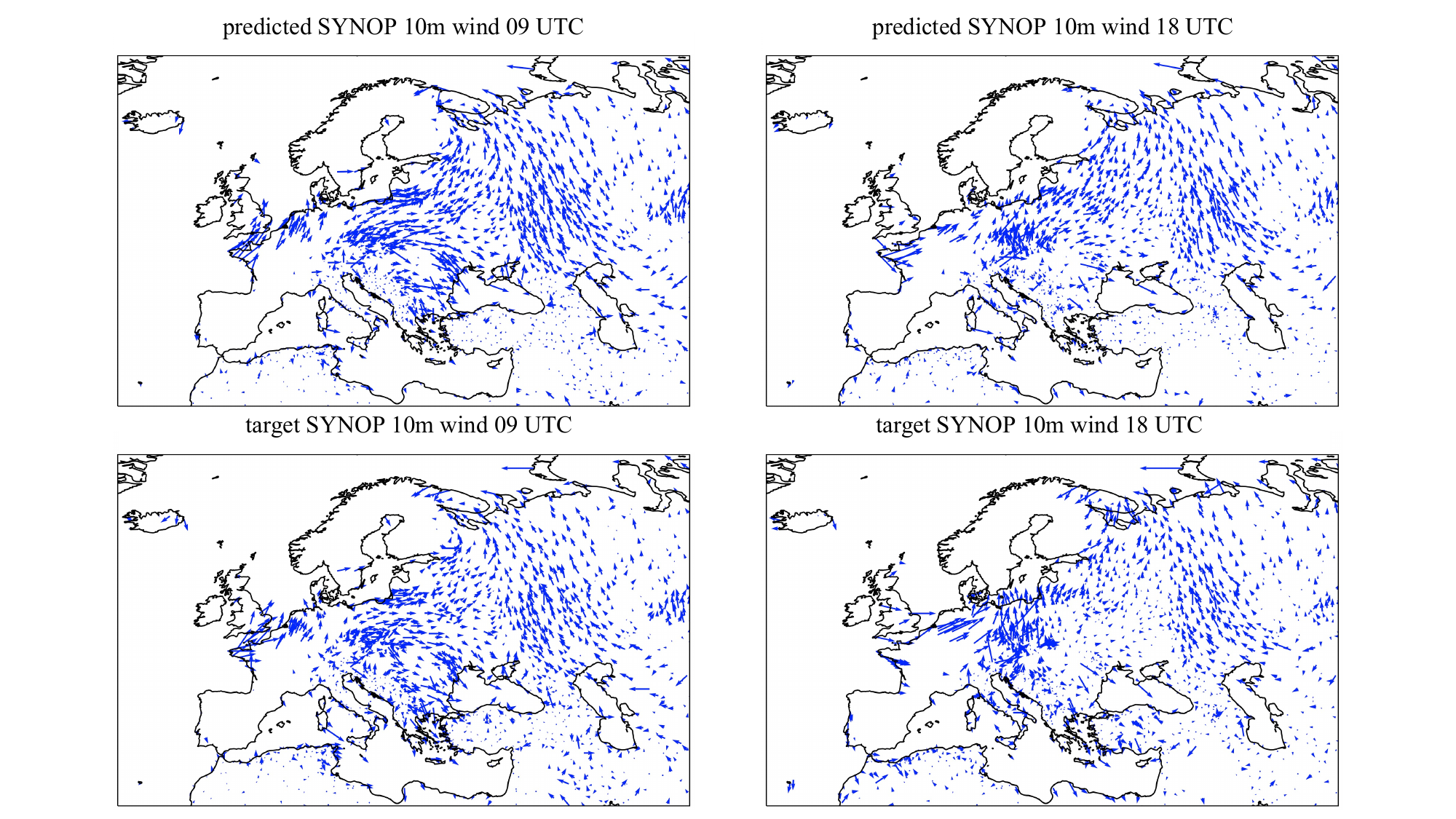}
    \caption{ Example of 10m winds inferred from satellite brightness temperatures alone. Predicted SYNOP 10m wind (upper panels) and verifying target SYNOP 10m wind (lower panels) at 9UTC and 18UTC for a case on February 18th 2022. }
    \label{fig:fig_5}
\end{figure}

Finally, we present evidence suggesting that the network can successfully produce plausible weather parameter forecasts exploiting the learned correlations between real SYNOP measurements and satellite observations. In Figure~\ref{fig:fig_5} we see an example of predicted 10m SYNOP winds (upper panels) compared to the real SYNOP wind observations (lower panels) in the target 12-hour window during an event on February 18th 2022 (a deep low-pressure area causing strong cyclonic winds over continental Europe and even an intense Medicane system in the Eastern Mediterranean causing strong winds on the North African coast). While there are clearly some differences in magnitude and direction between the forecast winds and the real observed wind vectors, the overall picture is a strikingly realistic representation of a rather complex synoptic situation. Furthermore, in this rapidly evolving synoptic situation, it is particularly impressive that the forecasted winds manage to follow the rapid changes in intensity and direction of the observed winds. 

It must be reiterated that SYNOP observations are diagnostic in these experiments, in that no real wind observations are provided as input to the forecast. The forecast winds in Figure~\ref{fig:fig_5} are completely diagnosed from the satellite radiances that were observed in the previous 12 hours, exploiting the correlations between historical SYNOP and satellite radiances that were learned during the network training. An examination of the input satellite radiances suggests that during this particular case, much of the wind information was being inferred from ATMS. The microwave brightness temperatures provide clear information on the strong thermal gradients and the passage of the main synoptic features through the region. 

\section*{Discussion and future perspectives}

Here we have proposed a new and radical approach to weather forecasting by training a neural network to perform skilful weather forecasts purely from observations with no influence of a physics-based model, analysis or reanalysis dataset. In conventional NWP, we go to extraordinary lengths to convert many millions of observations to gridded fields of unmeasured quantities, to initialise conventional forecasting models with an imperfect representation of physics. The influence of the observations upon the forecast quality effectively stops at the initial conditions and limits the information from observations that can be exploited for a forecast. In any time-lapse animation of satellite cloud imagery, one sees a graphic illustration of exactly how our atmosphere actually evolves in time. In this new approach we use observations to initialise a model of the atmosphere (in observation space) learned directly from the observations themselves. Unlike other recent studies (e.g. \cite{vaughan2024aardvark, huang2024diffda, xiao2024coupling}) there is no explicit data assimilation step. 

There are a number of examples of extremely interesting work reported in the literature where observations are predicted with machine learning, but we feel it is important to point out the differences to this approach. Originally using feature tracking and extrapolation (e.g. \cite{bowler2006steps}), more recent nowcasting studies have demonstrated success using machine learning techniques \citep{ravuri2021skilful, espeholt2022deep, zhang2023skilful}. However, these nowcasting applications only ingest and predict a single variable, and capturing only a small fraction of the complete atmospheric state restricts the effective forecast range to just a few hours ahead. \citet{andrychowicz2023deep} were able to extend the skilful forecast range up to a day ahead by augmenting the observational data with gridded NWP analysis fields which were used as both input and targets in the training. But here we use no NWP analysis data, instead including a far wider range of observational data sources to provide the network with richer information about the global temperature and humidity structure of the atmosphere.

The results presented here show that, even with a prototype neural network, realistic predictions out to 12-hours can be obtained using a rather limited set of satellite and in-situ training observations. We see evidence that the network is learning not just simple advection of weather systems, but also more complex evolutions in time and space. Finally, through learning robust relationships between different observation types, the approach is able to generate extremely realistic forecasts of useful weather parameters by predicting future weather parameter observations such as SYNOP winds and temperatures.

Of course, on the basis of these very preliminary results we cannot speculate upon how realistic longer-range predictions directly from observations might be. Established data driven models (such as the AIFS, GraphCast and Pangu-Weather) trained from reanalysis data have clearly demonstrated this longer-range capability so we have some reason for optimism. We are currently working towards this by building scalable and more sophisticated neural networks trained and initialised with all operationally used observations. 

If successful, a particularly exciting prospect of this approach is that it is readily (indeed immediately) extendable to forecasts of the full Earth system. In conventional NWP, completely different physics-based models for each Earth component are used (e.g. ocean and atmosphere), requiring completely separate data assimilation systems to support them, and complex coupling technology to allow communication between different components. In the fully data-driven approach presented here, the network is completely agnostic (like the observations themselves) to which Earth component the observations have come from. Providing they are in the same training data, the network is free to learn relationships between ocean measurements (such as Argo floats or drifters) and satellite measurements of the atmosphere and ocean surface (such as radiances, scatterometry and altimetry). By identifying and exploiting those correlations which improve forecast accuracy, direct observation prediction is implicitly and instantly capable of fully coupled forecasting. In a similar manner, observations of atmospheric composition (e.g. aerosol or greenhouse gas concentration) can also be simultaneously introduced into the training data, allowing the network to learn to predict the future evolution of these, exploiting spatial and time correlations with other (meteorological) observations in the training data (such as wind observations). 

Finally, we see opportunities for direct observation prediction to exploit additional observations that are currently not used by conventional data assimilation systems. For example, satellite radiance measurements in the visible part of the spectrum have such complex radiative transfer that they are not yet assimilated in most global NWP systems. Yet in any animation of visible imagery one can clearly see the movement of weather patterns around the globe, and it seems entirely plausible that the network would readily exploit this information in its predictions of weather parameters without needing a detailed physics-based knowledge of the observation origin. 

\section*{Appendix A: Glossary of satellite sensor acronyms }

\begin{table}[h!]
\noindent
    \begin{tabular}{ll}
AMSU &Advanced Microwave Sounding Unit (carried by NOAA and EUMETSAT satellites) \\
& \\
ATMS &Advanced Technology Microwave Sounder (carried by NOAA satellites)\\ 
& \\
IASI &Infrared Atmospheric Sounding Interferometer (carried by EUMETSAT satellites)\\
& \\
AVHRR &Advanced Very High-Resolution Radiometer (carried by NOAA and EUMETSAT satellites)\\
& \\
SYNOP &Surface Synoptic Observations (WMO code type for this measurement)\\
    \end{tabular}

\end{table}

\printbibliography

@article{andrychowicz2023deep,
  title={Deep learning for day forecasts from sparse observations},
  author={Andrychowicz, Marcin and Espeholt, Lasse and Li, Di and Merchant, Samier and Merose, Alexander and Zyda, Fred and Agrawal, Shreya and Kalchbrenner, Nal},
  journal={arXiv preprint arXiv:2306.06079},
  year={2023}
}

@article{bi2023accurate,
  title={Accurate medium-range global weather forecasting with 3D neural networks},
  author={Bi, Kaifeng and Xie, Lingxi and Zhang, Hengheng and Chen, Xin and Gu, Xiaotao and Tian, Qi},
  journal={Nature},
  volume={619},
  number={7970},
  pages={533--538},
  year={2023},
  publisher={Nature Publishing Group UK London}
}

@article{bowler2006steps,
  title={STEPS: A probabilistic precipitation forecasting scheme which merges an extrapolation nowcast with downscaled NWP},
  author={Bowler, Neill E and Pierce, Clive E and Seed, Alan W},
  journal={Quarterly Journal of the Royal Meteorological Society: A journal of the atmospheric sciences, applied meteorology and physical oceanography},
  volume={132},
  number={620},
  pages={2127--2155},
  year={2006},
  publisher={Wiley Online Library}
}

@article{espeholt2022deep,
  title={Deep learning for twelve hour precipitation forecasts},
  author={Espeholt, Lasse and Agrawal, Shreya and S{\o}nderby, Casper and Kumar, Manoj and Heek, Jonathan and Bromberg, Carla and Gazen, Cenk and Carver, Rob and Andrychowicz, Marcin and Hickey, Jason and others},
  journal={Nature communications},
  volume={13},
  number={1},
  pages={1--10},
  year={2022},
  publisher={Nature Publishing Group}
}

@article{lam2023learning,
  title={Learning skillful medium-range global weather forecasting},
  author={Lam, Remi and Sanchez-Gonzalez, Alvaro and Willson, Matthew and Wirnsberger, Peter and Fortunato, Meire and Alet, Ferran and Ravuri, Suman and Ewalds, Timo and Eaton-Rosen, Zach and Hu, Weihua and others},
  journal={Science},
  volume={382},
  number={6677},
  pages={1416--1421},
  year={2023},
  publisher={American Association for the Advancement of Science}
}

@article{lang2024aifs,
  title={AIFS-ECMWF's data-driven forecasting system},
  author={Lang, Simon and Alexe, Mihai and Chantry, Matthew and Dramsch, Jesper and Pinault, Florian and Raoult, Baudouin and Clare, Mariana CA and Lessig, Christian and Maier-Gerber, Michael and Magnusson, Linus and others},
  journal={arXiv preprint arXiv:2406.01465},
  year={2024}
}

@article{price2023gencast,
  title={GenCast: Diffusion-based ensemble forecasting for medium-range weather},
  author={Price, Ilan and Sanchez-Gonzalez, Alvaro and Alet, Ferran and Ewalds, Timo and El-Kadi, Andrew and Stott, Jacklynn and Mohamed, Shakir and Battaglia, Peter and Lam, Remi and Willson, Matthew},
  journal={arXiv preprint arXiv:2312.15796},
  year={2023}
}

@article{ravuri2021skilful,
  title={Skilful precipitation nowcasting using deep generative models of radar},
  author={Ravuri, Suman and Lenc, Karel and Willson, Matthew and Kangin, Dmitry and Lam, Remi and Mirowski, Piotr and Fitzsimons, Megan and Athanassiadou, Maria and Kashem, Sheleem and Madge, Sam and others},
  journal={Nature},
  volume={597},
  number={7878},
  pages={672--677},
  year={2021},
  publisher={Nature Publishing Group UK London}
}

@article{rodgers1976retrieval,
  title={Retrieval of atmospheric temperature and composition from remote measurements of thermal radiation},
  author={Rodgers, Clive D},
  journal={Reviews of Geophysics},
  volume={14},
  number={4},
  pages={609--624},
  year={1976},
  publisher={Wiley Online Library}
}

@techreport{xiao2024coupling,
  title={Coupling the Data-driven Weather Forecasting Model with 4D Variational Assimilation},
  author={Xiao, Yi and Bai, Lei and Xue, Wei and Chen, Kang and Han, Tao and Ouyang, Wanli},
  year={2024},
  institution={Copernicus Meetings}
}

@article{zhang2023skilful,
  title={Skilful nowcasting of extreme precipitation with NowcastNet},
  author={Zhang, Yuchen and Long, Mingsheng and Chen, Kaiyuan and Xing, Lanxiang and Jin, Ronghua and Jordan, Michael I and Wang, Jianmin},
  journal={Nature},
  volume={619},
  number={7970},
  pages={526--532},
  year={2023},
  publisher={Nature Publishing Group UK London}
}

@article{huang2024diffda,
  title={Diffda: a diffusion model for weather-scale data assimilation},
  author={Huang, Langwen and Gianinazzi, Lukas and Yu, Yuejiang and Dueben, Peter D and Hoefler, Torsten},
  journal={arXiv preprint arXiv:2401.05932},
  year={2024}
}

@article{vaughan2024aardvark,
  title={Aardvark Weather: end-to-end data-driven weather forecasting},
  author={Vaughan, Anna and Markou, Stratis and Tebbutt, Will and Requeima, James and Bruinsma, Wessel P and Andersson, Tom R and Herzog, Michael and Lane, Nicholas D and Hosking, J Scott and Turner, Richard E},
  journal={arXiv preprint arXiv:2404.00411},
  year={2024}
}

@article{bauer2015quiet,
  title={The quiet revolution of numerical weather prediction},
  author={Bauer, Peter and Thorpe, Alan and Brunet, Gilbert},
  journal={Nature},
  volume={525},
  number={7567},
  pages={47--55},
  year={2015},
  publisher={Nature Publishing Group UK London}
}

@article{vaswani2017attention,
  title={Attention is all you need},
  author={Vaswani, Ashish and Shazeer, Noam and Parmar, Niki and Uszkoreit, Jakob and Jones, Llion and Gomez, Aidan N and Kaiser, {\L}ukasz and Polosukhin, Illia},
  journal={Advances in neural information processing systems},
  volume={30},
  year={2017}
}

@article{rabier20004dvar,
author = {Rabier, F. and Järvinen, H. and Klinker, E. and Mahfouf, J.-F. and Simmons, A.},
title = {The ECMWF operational implementation of four-dimensional variational assimilation. I: Experimental results with simplified physics},
journal = {Quarterly Journal of the Royal Meteorological Society},
volume = {126},
number = {564},
pages = {1143-1170},
doi = {https://doi.org/10.1002/qj.49712656415},
url = {https://rmets.onlinelibrary.wiley.com/doi/abs/10.1002/qj.49712656415},
eprint = {https://rmets.onlinelibrary.wiley.com/doi/pdf/10.1002/qj.49712656415},
year = {2000}
}

@article{hersbach2020era5,
author = {Hersbach, Hans and Bell, Bill and Berrisford, Paul and Hirahara, Shoji and Horányi, András and Muñoz-Sabater, Joaquín and Nicolas, Julien and Peubey, Carole and Radu, Raluca and Schepers, Dinand and Simmons, Adrian and Soci, Cornel and Abdalla, Saleh and Abellan, Xavier and Balsamo, Gianpaolo and Bechtold, Peter and Biavati, Gionata and Bidlot, Jean and Bonavita, Massimo and De Chiara, Giovanna and Dahlgren, Per and Dee, Dick and Diamantakis, Michail and Dragani, Rossana and Flemming, Johannes and Forbes, Richard and Fuentes, Manuel and Geer, Alan and Haimberger, Leo and Healy, Sean and Hogan, Robin J. and Hólm, Elías and Janisková, Marta and Keeley, Sarah and Laloyaux, Patrick and Lopez, Philippe and Lupu, Cristina and Radnoti, Gabor and de Rosnay, Patricia and Rozum, Iryna and Vamborg, Freja and Villaume, Sebastien and Thépaut, Jean-Noël},
title = {The ERA5 global reanalysis},
journal = {Quarterly Journal of the Royal Meteorological Society},
volume = {146},
number = {730},
pages = {1999-2049},
doi = {https://doi.org/10.1002/qj.3803},
url = {https://rmets.onlinelibrary.wiley.com/doi/abs/10.1002/qj.3803},
eprint = {https://rmets.onlinelibrary.wiley.com/doi/pdf/10.1002/qj.3803},
year = {2020}
}

\end{document}